\documentclass[manuscript]{aastex}

\shorttitle{Probing Brownstein--Moffat Gravity via Numerical Simulations}
\shortauthors{Brandao \& de Araujo}

\begin{document}
\title{Probing Brownstein--Moffat Gravity via Numerical Simulations}

\author{C. S. S. Brandao
and J. C. N. de Araujo}
\affil{Divis\~{a}o de Astrof\'{i}sica, Instituto Nacional de Pesquisas Espaciais, \\
S. J. Campos, SP 12227-010, Brazil}
\email{claudio@das.inpe.br, jcarlos@das.inpe.br}

\begin{abstract}
In the standard scenario of the Newtonian gravity, a late-type galaxy (i.e., a spiral galaxy) is well described by a disk and a bulge embedded in a halo mainly composed by dark matter. In Brownstein-Moffat gravity, there is a claim that late-type galaxy systems would not need to have halos, avoiding as a result the dark matter problem, i.e., a modified gravity (non-Newtonian) would account for the galactic structure with no need of dark matter. In the present paper, we probe this claim via numerical simulations. Instead of using a ``static galaxy," where the centrifugal equilibrium is usually adopted, we probe the Brownstein-Moffat gravity dynamically via numerical $N$-body simulations.
\end{abstract}




\section{Introduction}

Spiral galaxies are very common in the universe. When studied with spectroscopic methods, these beautiful objects usually reveal to rotate faster than it would do if only their visible matter were taken into account. This is the missing mass problem. Then, one generally believes that there is a dark matter component surrounding their disks, making them to rotate in that way. In other words, the canonical model says that spiral galaxy rotation curves inferred from observations could be accounted for the dark matter halos that embed the galaxies \cite[see, e.g.,][]{rubin,bahcall}.  Therefore, this additional distribution of mass causes an additional acceleration and the rotation curves can be accounted for.

This picture to the understanding of the structure of late-type morphologies finds support in the present canonical physics, that considers the general relativity the most reliable theory of gravitation, and, as a consequence, due to the correspondence principle, the Newtonian gravitation at the weak field limit.

However, in order to avoid the dark matter problem, many authors consider modifications in the Newtonian gravity at large scales, in other words, they consider that, in galactic or even cosmological scales, the law of gravity is different from the Newtonian one. In this way, contrary to the dark matter paradigm, these authors argue that there is only a disk, but due to another kind of law of gravity  at large scales, it could well rotate in accord with the observed rotation curves \cite[see, e.g., ][ and references therein]{milgrom1,milgrom2,moffat1,moffat2,bm2006a}.

It is worth mentioning that \cite{bm2006b} also argue that it is possible to explain the X-ray cluster masses without dark matter using this very theory.

In particular, the present work is concerned with the study made by \citet[][ hereafter BM]{bm2006a}, who present an alternative approach to explain the rotation curves of spiral galaxies, using an alternative theory of gravitation studied by \citet[][ hereafter Moffatian gravity]{moffat1}. Our main aim is to probe this theory using numerical simulations.

The Moffatian gravity is a covariant theory of gravity based on the coupling of the Einsteinian gravity with a massive skew symmetric field $F_{\alpha \beta \gamma}$. Here, we do not present and discuss  this alternative theory of gravity in detail, since we are only concerned with its weak field limit.

BM claim that in the weak field limit the gravitational potential of a point source of mass $M$ reads

\begin{equation}
\Phi(r) = - \frac{GM}{r} \left\{ 1 + \sqrt{\frac{M_0}{M}} \left[ 1 - \exp  \left(- \frac{r}{r_0} \right) \right] \right\} ,
\label{moffat_equation}
\end{equation}

\par\noindent where $G$ is the Newtonian constant of gravity, $r$ is the distance to the source, $M_0$ is a coupling parameter and $r_0$ is a characteristic length. From the BM potential, the gravitational acceleration by a particle of mass $M$ reads

\begin{equation}
a(r) = - \frac{GM}{r} \left\{ 1 + \sqrt{\frac{M_0}{M}} \left[ 1 - \exp  \left(- \frac{r}{r_0} \right) \left( 1 +  \frac{r}{r_0} \right)   \right] \right\}.
\label{moffat_aceleration}
\end{equation}

BM state that the above equations hold in any environment. To apply these equations to model their rotation curves,
they consider the following procedure: they integrate these equations for a spherically symmetric system. As a result, they obtain the following equations:

\begin{equation}
\Phi(r) = - \frac{GM(r)}{r} \left\{ 1 + \sqrt{\frac{M_0}{M}} \left[ 1 - \exp  \left(- \frac{r}{r_0} \right) \right] \right\} ,
\label{moffat_equation2}
\end{equation}

\par\noindent and

\begin{equation}
a(r) = - \frac{GM(r)}{r} \left\{ 1 + \sqrt{\frac{M_0}{M}} \left[ 1 - \exp  \left(- \frac{r}{r_0} \right) \left( 1 +  \frac{r}{r_0} \right)   \right] \right\}.
\label{moffat_aceleration2}
\end{equation}

Now, the $M$ outside the braces becomes $M(r)$, but the $M$ inside the braces does not change. Therefore, for given values of $r_0$, $M_0$ and $M$ (the total mass of the system) the expression inside the braces depends only on $r$.
Assuming centrifugal equilibrium and a given mass distribution, i.e. $M(r)$, BM model their rotation curves.

Also, note that Equation (\ref{moffat_aceleration2}) obeys the superposition principle, which is an essential property if one intends to make use of $N$-body simulations.

We note from the above equations that the Moffatian gravity can be adjusted to a large range of lengths and masses, by means of the parameters $r_0$ and $M_0$, that depend on the system under investigation. Hence, they could, in principle, explain many physical effects that the canonical theory attributes to the dark matter: galaxy rotation curves, galactic cluster masses and so on, because the BM's parameters depend on the total mass of the system under study.

BM apply their modified gravity to explain the rotation curves of a sample of observed spiral galaxies and yield the following set of parameters, namely, $M_0 = 96 \times 10^{10} M_{\odot}$ and $r_0 = 13.96 $ kpc \citep{bm2006a}. As usual in this kind of study, they consider centrifugal equilibrium, which means in the end that their galaxy model is ``static."

It is well known, from the $N$-body simulation studies, that disks built with exponential-Spitzer laws (hereafter SdMH disks), when submitted to the Newtonian potentials, reveal to have secular centrifugal equilibrium, maintaining their density profiles for gigayears of simulated time \citep{sw1999,sdmh2005}. However, under the BM's picture, we do not know how long the BM potential can maintain galactic disks in dynamical equilibrium, and in fact if it really can. This is the main aim of the present paper.

In previous works \citep{ca2009a, ca2009b,ca2009c}, we developed an $N$-body method to study alternative theories of gravity at galactic scales. In particular, we have studied an Yukawian gravitational potential and have shown that this potential is viable only if the Yukawian parameter is such that the Yukawian potential is nearly Newtonian.

In the present work, we also use $N$-body simulations to construct and evolve spiral galaxies now submitted to the potential and acceleration given by Equations (\ref{moffat_equation2}) and (\ref{moffat_aceleration2}), respectively. We adopted, in these simulations, the same set of fitted parameters ($M_0 = 96 \times 10^{10} M_{\odot}$ and $r_0 = 13.96 $ kpc) as by BM.

Our aim is to probe if the BM model is dynamically consistent, as it should be in order to be considered a realistic model. In this way, we are able to test the BM's claim and verify the reliability of this kind of law of gravity, using the concepts and techniques of the galactic dynamics studies \citep{bt2008}.

In Section \ref{numtech}, we present the $N$-body code used in this work and the numerical techniques used to model spiral galaxies with the BM potential; in Section \ref{simulresul}, the results of the numerical simulation are present; and finally, in Section \ref{conclusao}, we discuss our main findings and conclusions.

\section{Numerical Techniques}
\label{numtech}

\subsection{The Modified Code}
\label{test1}

We performed the simulations with a modified and tested version of the Gadget-2 code \citep{springel2005}. As discussed in our previous papers \citep{ca2009a,ca2009b,ca2009c}, this choice is based on the collisionless nature of galaxies modeled by particles, and the method used by Gadget-2 to build the recursive tree in order to compute forces and potentials.

We have modified the Gadget-2 and replaced the Newtonian potential and acceleration by the expressions given by Equations (\ref{moffat_equation2}) and (\ref{moffat_aceleration2}) just in the respective code's instructions. We have extensively  tested the code efficiency to calculate potentials via the tree method, using a typical initial snapshot composed by a disk embedded in a dark matter halo.

It is worth noting that we included a halo only to test the code accuracy, since in the probes of the ability of the BM gravity to generate spiral galaxies with flat rotation curves, without dark matter, the halo is not included in the simulations.

The Newtonian and the Moffatian codes were set up with the tolerance parameter $\theta=0.8$, maximizing the tree code's performance and softening length $l_{\rm{dh}} = 0.33$ kpc for the dark halo and $l_{\rm{d}} = 0.15$ kpc for the baryonic disk. This set of softening lengths yields conservation of the total energy better than other sets. The procedure to choose the values for  $\theta$, $l_{\rm{dh}}$ and $l_{\rm{d}}$ adopted here is analog to that used in our previous papers \citep[][ and references therein]{ca2009a}.

\subsection{Initial Conditions}

We have studied three models: model I a typical exponential-Spitzer disk with a dark matter halo (Newtonian galaxy); model II a second one, derived from the first one, where the system is built and the halo is removed before running the simulation with a Moffatian potential (pseudo-Moffatian galaxy); and model III a pure BM's disk model (pure Moffatian galaxy). Below, we give the reasons why such models are chosen.

Model I follows the prescriptions described by  \citet[][ and references therein]{sdmh2005}, but with the difference that we are now including neither bulges nor gas particles, because we do not need a detailed Newtonian model, since our main aim is to investigate the dynamical properties of disk galaxies under alternative potentials.  The reason for this choice has to do with their dynamical stability and computational simplicity to model them, as explained in detail in \citet{sdmh2005}. Besides, it is worth stressing that SdMH disks reproduce all observational features of late-type systems.

In one of our previous papers \citep{ca2009c}, we described the principal recipes to model such galaxies. Besides
the parameters discussed in the previous section, the models are composed by $N_{\rm{halo}} = 30,000$ particles for the dark matter halo and $N_{\rm{disk}} = 30,000$ for the disk. All these figures yield a conservation of energy that turns our galaxy model reliable, as the reader will find later on in this paper. Model I is then a Newtonian model for a spiral galaxy, which is consistent with observations, that we use to compare with the Moffatian model for spiral galaxies.

Model II is the same as model I but without dark halo particles, i.e., it is just the Model I from it we have extracted the halo particles. It is worth stressing that the halo potential is computed to make the rotation curve of the disk of model II as reliable as possible--like the observed ones.

After extracting the halo, the disk should be stable under the BM's potential.  This procedure is made to test the BM's hypothesis, which claims that disks can exist under Moffat's potential without halos, explaining rotation curves without exotic non-baryonic dark matter. In this way, we are able to test this hypothesis over a secular evolution of a disk.

Model III is similar to model II, but with the following differences: (1) we do not compute the halo's potential and (2) we replace the Newtonian potential originated from the disk \cite[see, e.g.][ Equation (2.170)]{bt2008} by the BM's potential originated from  all the positions of disk particles, using the BM's parameters $M_0 = 96 \times 10^{10} M_{\odot}$ and $r_0 = 13.96 $ kpc, as  fitted by \citet{bm2006a}.

In other words, we first build the positions of the particles and then compute their potentials using Equation (\ref{moffat_equation2}) over the mesh, where the velocities will be calculated \cite[see, e.g.,][]{ca2009c}. We use these results to integrate the Jean's equations, and then calculate the velocity dispersions. With this procedure, the initial disk is put in equilibrium with the BM's potential during the first snapshots of simulated time. We have set $N_{\rm{disk}} = 30,000$ particles to this disk (hereafter Moffatian disk or BM's disk for short) in order to maintain the same resolution set to the Newtonian model. Note that the disk thus modeled is not necessarily consistent with observations, as the reader will see later on.

The following set of parameters is chosen to the SdMH disks (model I)): the total mass ${M_{\rm{t}} = v^2_{200}/(10GH_0) = 0.98 \times 10^{12} M_{\odot}}$, where $v_{200} = 160\, \rm{km\, s^{-1}}$ is the virial velocity, $G$ is the gravitational constant, and $H_0 = 100\, \rm{km\, s^{-1}\, Mpc^{-1}}$, is the Hubble constant; the total mass of the disk $M_{\rm{disk}} = m_{\rm{d}} M_{\rm{t}}$, where $m_{\rm{d}}=0.041$ is a dimensionless fraction of the total mass; the disk scale length $h = 2.74$ kpc; the disk vertical scale height $z_0 \sim 0.2 h$; and the spin parameter $\lambda = 0.033$. We emphasize that this model has mass comparable to the Milk Way.

For model II we also use the same set of parameters used in model I. Recall, however, that after building the disk--halo system, we then extract the halo.

In the BM's disks, we maintain the total disk mass and the exponential-Spitzer profiles.  We built the disk particles from the density profiles, and used the BM's parameters  ($M_0 = 96 \times 10^{10} M_{\odot}$ and $r_0 = 13.96 $ kpc).

In the next section, we present the results of our simulations.

\section{Simulations and Results}
\label{simulresul}

\subsection{The Newtonian Model}

Although the Newtonian model can be considered as a canonical one by the current theory of galactic astronomy \citep{bm1998} and galactic dynamics \citep{bt2008}, it is important to bear in mind that here the Newtonian model is not the main result of our work. We have run the Newtonian simulations to compare it with models II and III. As it is well known from simulations of canonical late-type systems \citep{sdmh2005,ca2009c}, many observational features, as for example, the exponential radial density profile, the rotation curves, etc can be accounted for Newtonian models.

\begin{figure}
\epsscale{1.0}
\plotone{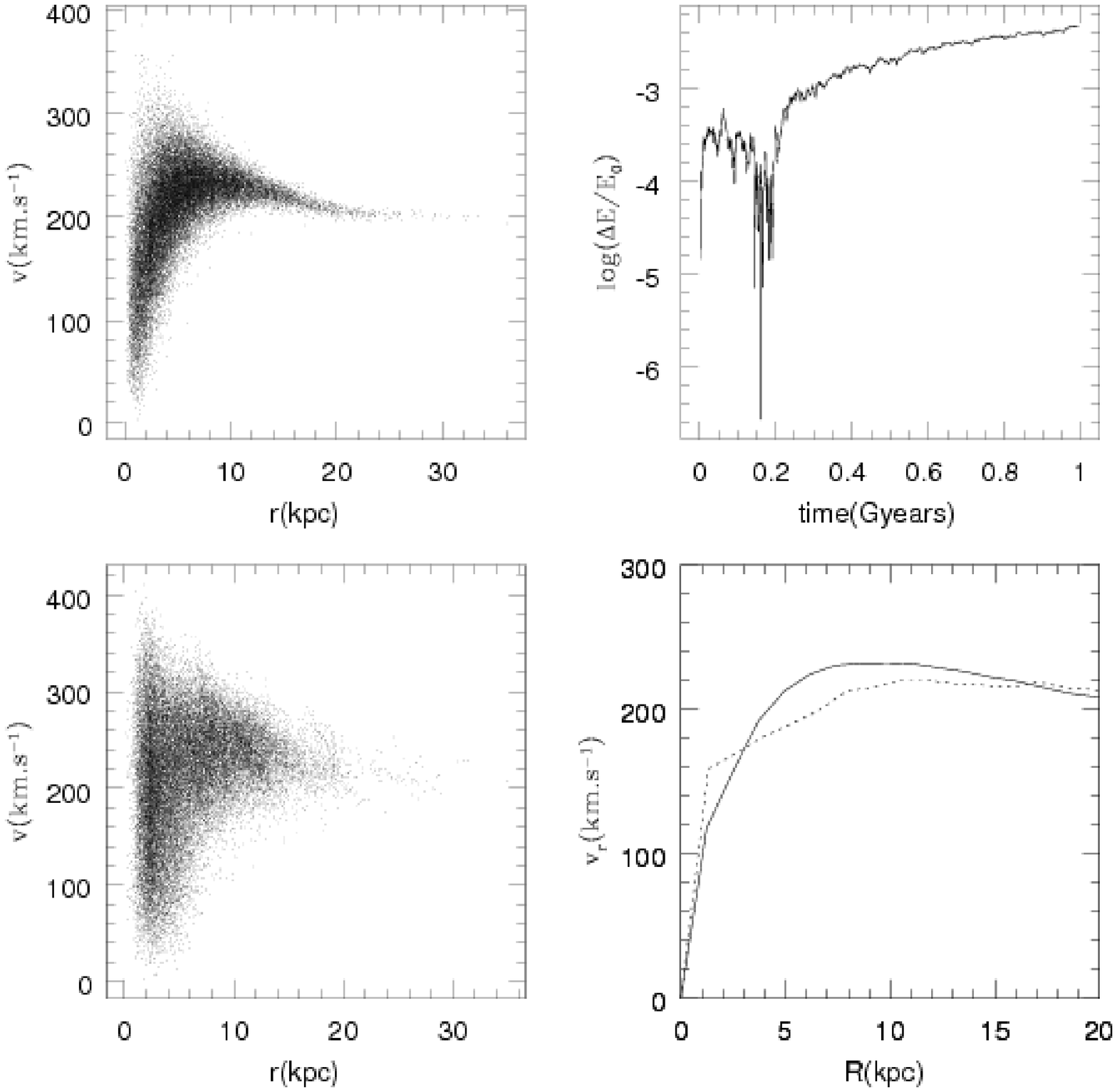}
\caption{Top left: phase space for initial snapshot data. Top right: energy conservation of the simulation. Bottom left: phase space for final snapshot data at 1 Gyr. Bottom right: rotation curves for initial (solid line) and final (dashed line) snapshots. \label{fig1}}
\end{figure}

\begin{figure}
\epsscale{1.0}
\plotone{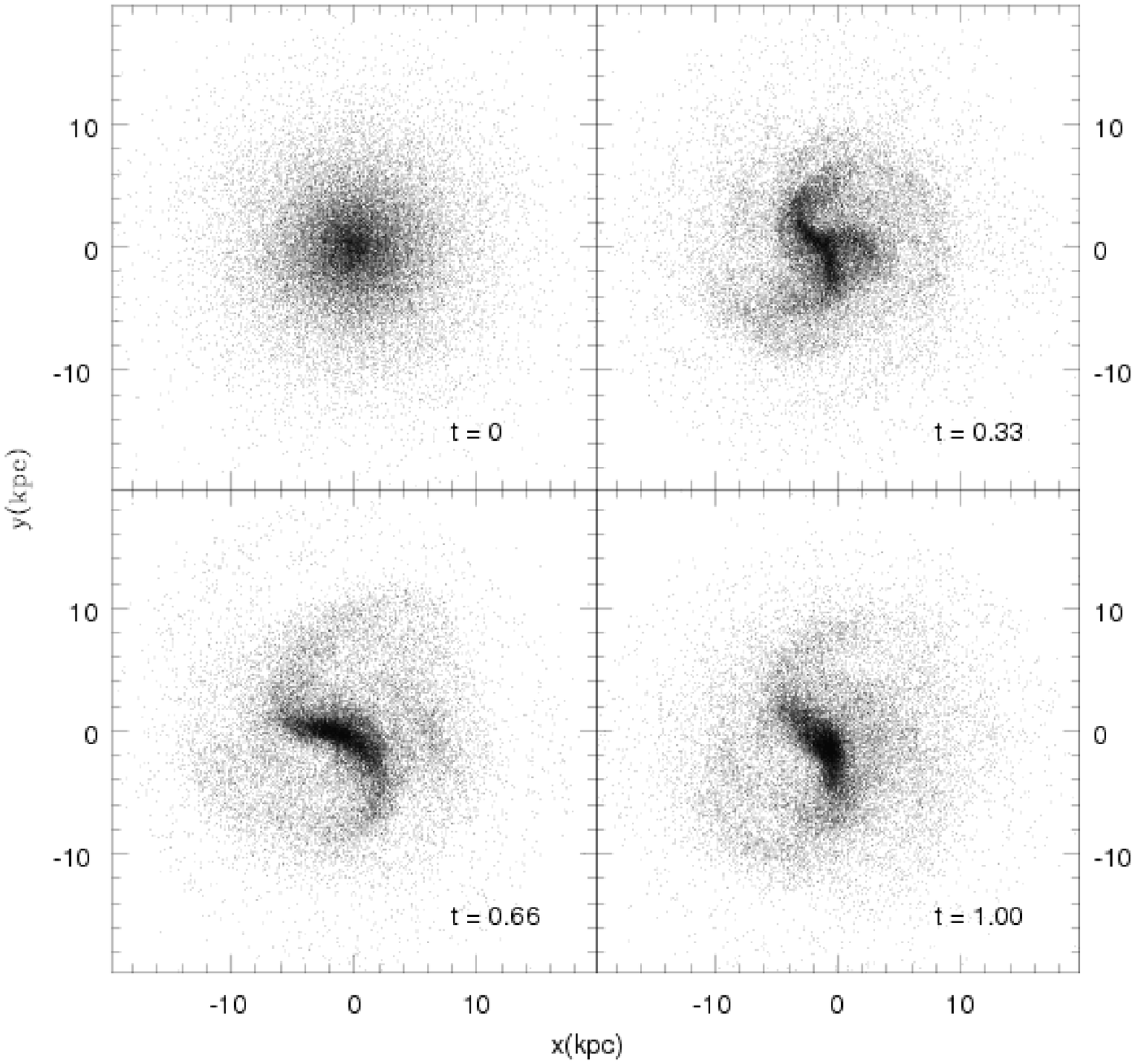}
\caption{Newtonian disk at $z$-projection at 0, 0.33, 0.66, and 1 Gyr of simulated time (indicated in the respective boxes). \label{fig2}}
\end{figure}

\begin{figure}
\epsscale{1.0}
\plotone{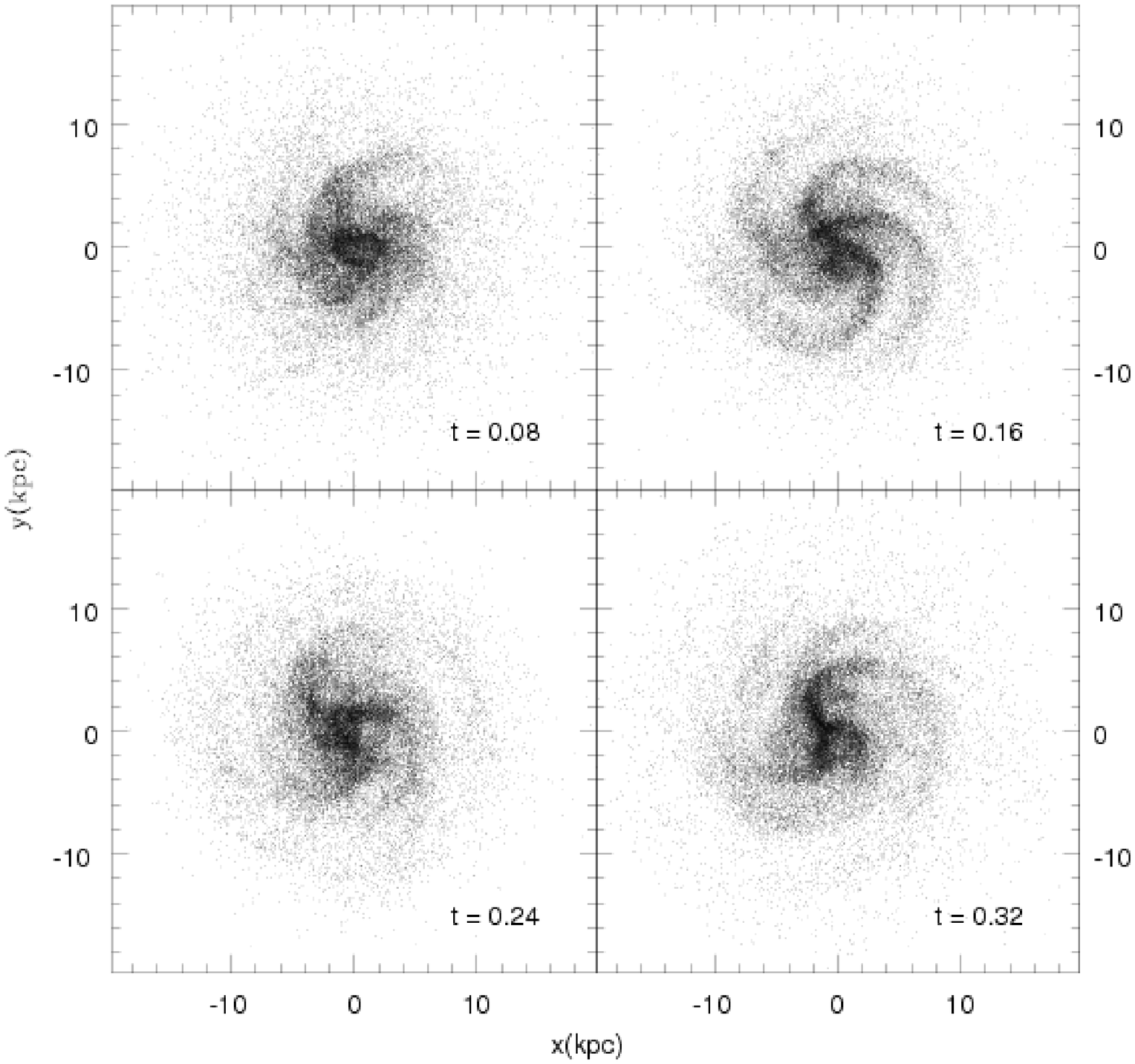}
\caption{First 300 Myr of simulated time to the Newtonian disk at $z$-projection. Time is indicated in the respective boxes.  \label{fig3}}
\end{figure}

In Figures (\ref{fig1})--(\ref{fig3}), we display the principal results of the Newtonian model we simulated. Figure \ref{fig1} displays a picture with four boxes: at the top left, phase space points $r \times v$ of the initial snapshot, where $r$ is the total distance from origin (0,0,0) in kpc and $v$ is the modulus of the velocity in $\rm{km\, s^{-1}}$; at the top right, the relative energy conservation is $\Delta E / E_0$, where $\Delta E = E(t)-E(0)$ and $E_0 \equiv E(0)$, where $E(t)$ denotes the total energy of the system at time $t$; at bottom left, the phase space points at final time $t = 1$ Gyr; and at the bottom right, the rotation curve is $R \times v_r$, where R is the radial cylindrical coordinate and $v_r$ is the rotation velocity. Note the similarity between the initial and the 1 Gyr rotation curves; one concludes that early-type galaxy structure can be accounted for the Newtonian model quite well, as is well known. Later on we will see that the comparison between the density profiles at these same different times corroborates such a conclusion.

Figures \ref{fig2} and \ref{fig3} show the particle's positions in the $z$-projection at respective times indicated in the boxes. We note from these figures that the simulation follows a typical behavior with the formation
of spiral arms and central bar. These features are expected from simulations of late-type morphologies, as explained
by \citet[][ and references therein]{ca2009c}. In the first 0.33 Gyr, the disk has spiral arms and the system evolves to a central bar with two spiral arms afterward. In other words, this simulation is reliable and represents a standard typical late-type system without external influences. Energy conservation is better than $0.1 \%$.

\subsection{Moffatian Models}

We show above that the Newtonian disks maintain their physical properties for many crossing times.  Now, when a spiral galaxy is submitted to the BM potential, an anomalous behavior occurs to the disk's structure, as we will see in this section.

\begin{figure}
\epsscale{1.0}
\plotone{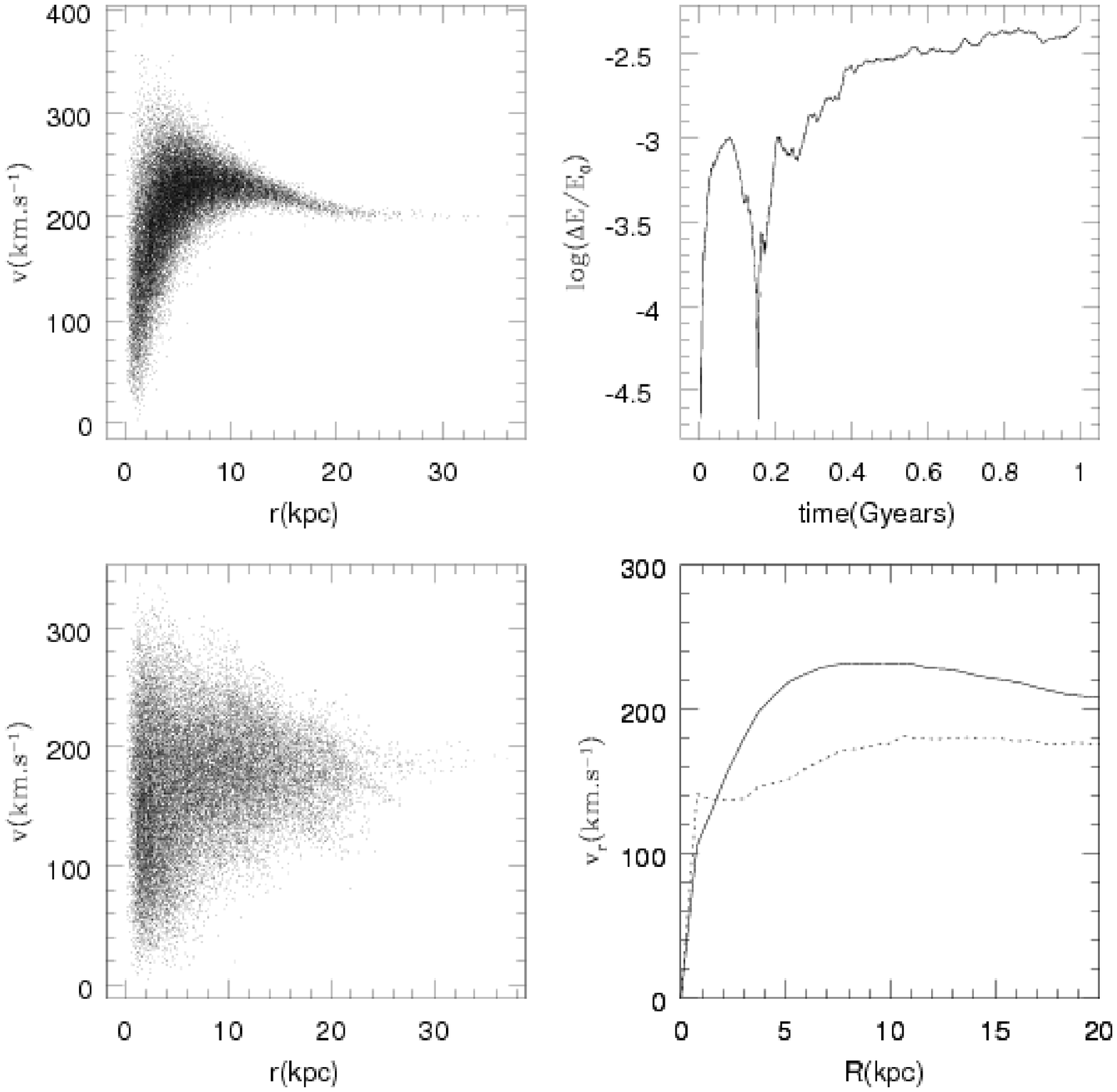}
\caption{Top left: phase space for initial snapshot data. Top right: energy conservation of the simulation. Bottom left: phase space for final snapshot data at 1 Gyr. Bottom right: rotation curves for initial (solid line) and final (dashed line) snapshots. \label{fig4}}
\end{figure}

In Figure \ref{fig4}, we display the same as in Figure \ref{fig1}.  In the top right panel of Figure \ref{fig4}, we display the evolution of the energy violation. Note that after 1 Gyr (end of our simulation), the violation is less than 1$\%$, showing the quality of our simulation.

From this simulation, we also see that the particles in the phase space are more scattered as compared to the Newtonian simulation. This scattering is interpreted as a loss of information from the initial configuration. Note, at the bottom right panel, that the disk rotates slower than in its initial time, even though it still resembles a typical rotation curve. At first glance, it seems, by only examining Figure \ref{fig4}, that the Moffatian potential can really explain BM's disks, and no dark matter is needed to maintain secularly stable exponential disks.

\begin{figure}
\epsscale{1.0}
\plotone{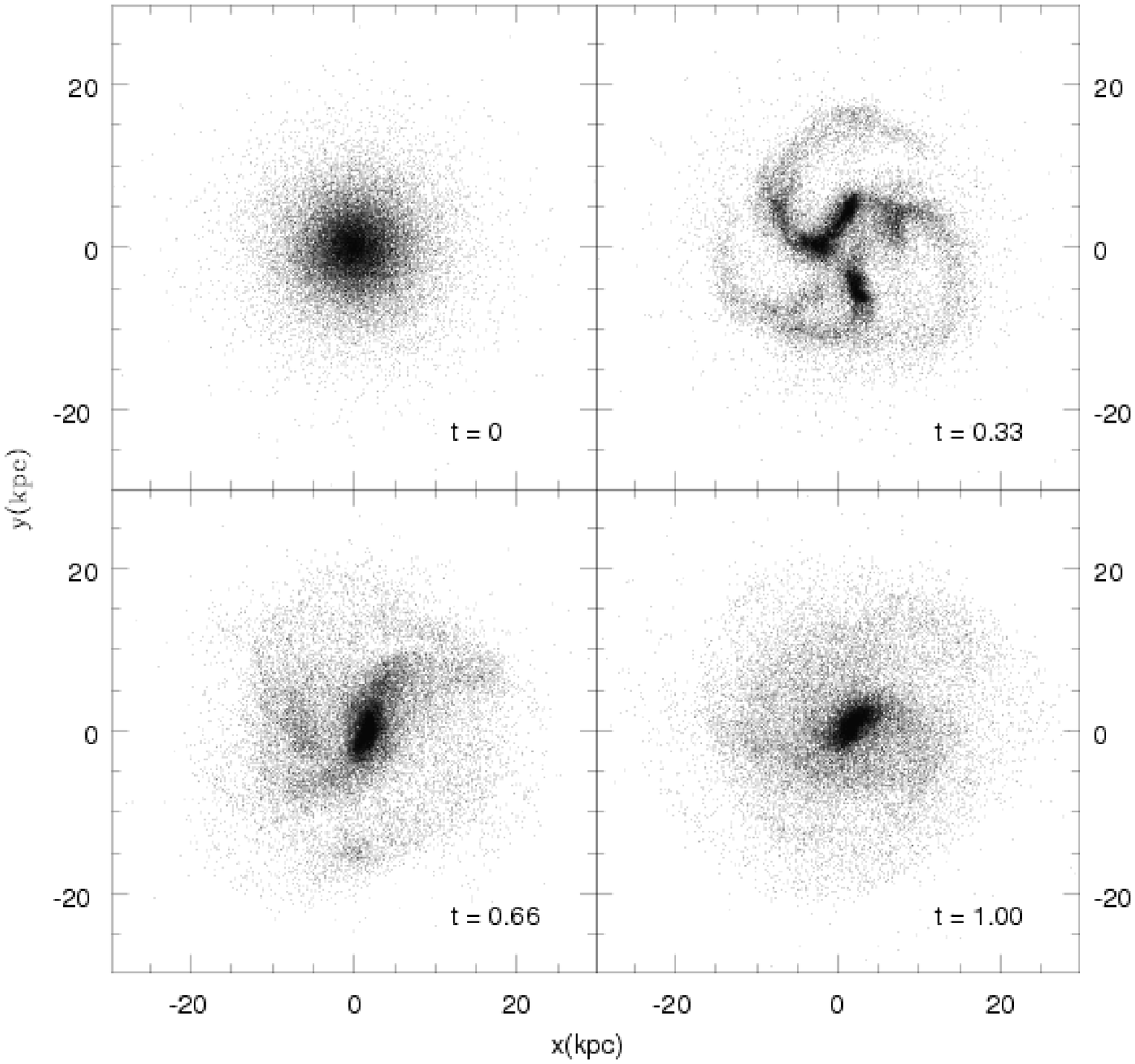}
\caption{Pseudo-Newtonian disk at $z$-projection at 0, 0.33, 0.66 and 1 Gyr of simulated time (indicated in the respective boxes). \label{fig5}}
\end{figure}

\begin{figure}
\epsscale{1.0}
\plotone{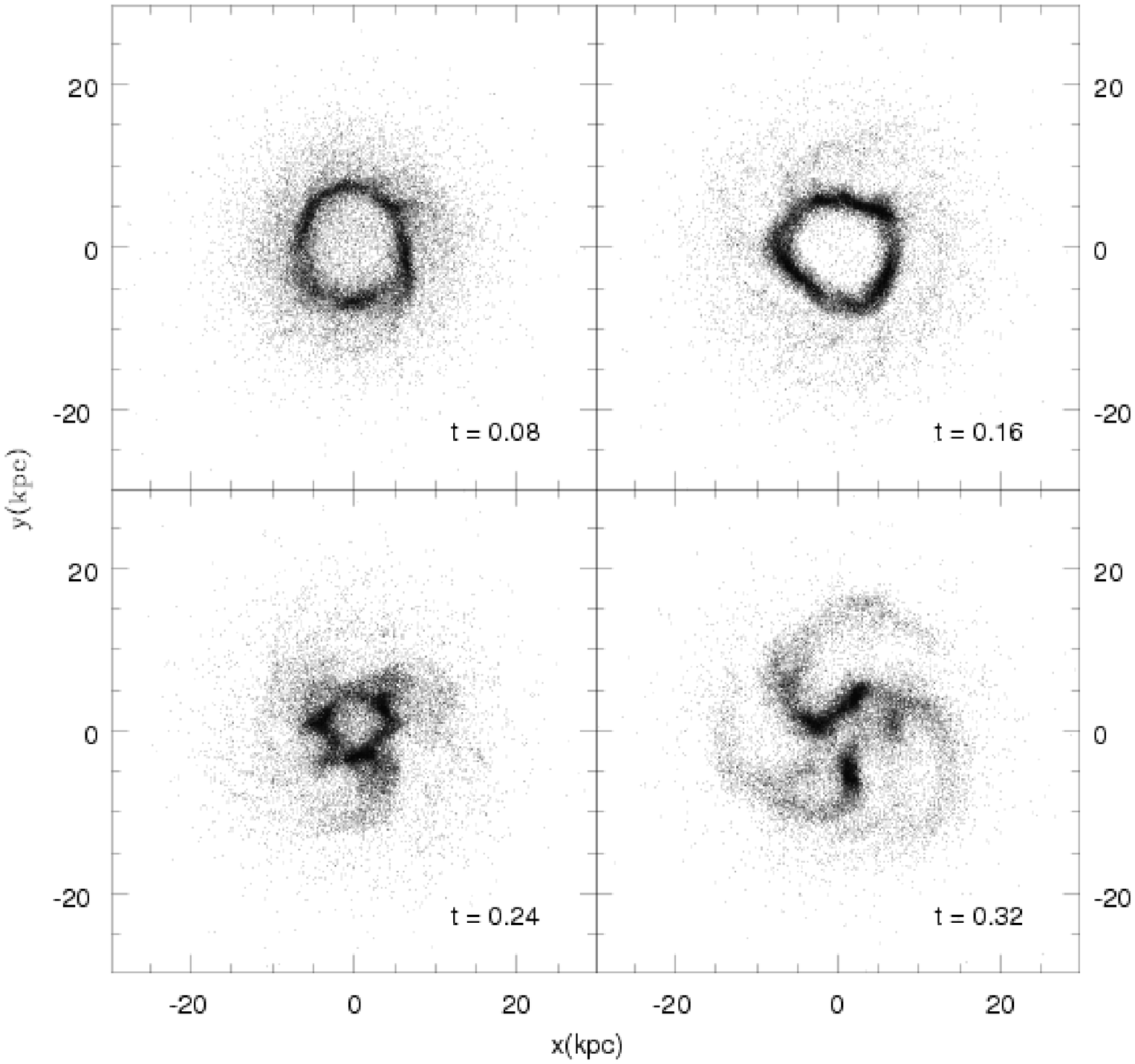}
\caption{First 300 Myr of simulated time to the pseudo-Newtonian disk at $z$-projection. Time is indicated in the respective boxes. \label{fig6}}
\end{figure}

We display snapshots of the disk's particles in Figures \ref{fig5} and \ref{fig6}. Examining these snapshots, we see that at 0.08, 0.16, 0.24, and 0.32 Gyr times the system presents atypical substructures, in particular in its central region ($R \lesssim 10.0$ kpc). As the disk evolves, an annular substructure appears, and the central region becomes almost empty of particles, a void. Subsequently, the big ring collapses and fragments into three spiral unbound arms. At this stage, there is no core.

At $t = 0.66$ Gyr, the arms merge themselves and become a bar-like structure surrounded by two weak spiral arms. This bar-like structure is small when compared to the Newtonian case. This anomalous behavior is due to the fact that the particles are not in perfect initial equilibrium with the Moffatian potential.

However, it is important to bear in mind that we are testing the BM claim, which considers that the Moffatian potential can explain many observed rotation curves of galactic disks without dark matter halos. We recall that our simulated initial disk reproduces the observational features of real galaxies: exponential-Spitzer disk, velocity dispersions, rotation curves and baryonic mass, and has secular equilibrium in the Newtonian case. From this point of view, we would expect that, if BM hypotheses were reliable, our disk would be stable all the time in the simulation. We obtain that an equilibrium configuration is reached only after $t \gtrsim 0.66$ Gyr. After this time, the system is similar, in some dynamical aspects, to the Newtonian case, but we will see below that this similarity is only apparent.

The results of model II's simulations bring some interesting questions. One could claim that the simulations of model II need a fine tuning of the parameters $\rm{M_0}$ and $\rm{r_0}$ to maintain the entire disk in equilibrium. These parameters could, therefore, be changed to avoid atypical rings and central voids during the first 0.33 Gyr. To study these sub-morphology features, we have made some tests changing the parameters $M_0$ and $r_0$, but we have not found any set of parameters that left the simulations just like the Newtonian one, which, it is worth stressing again, is consistent with observations.

\begin{figure}
\epsscale{1.0}
\plotone{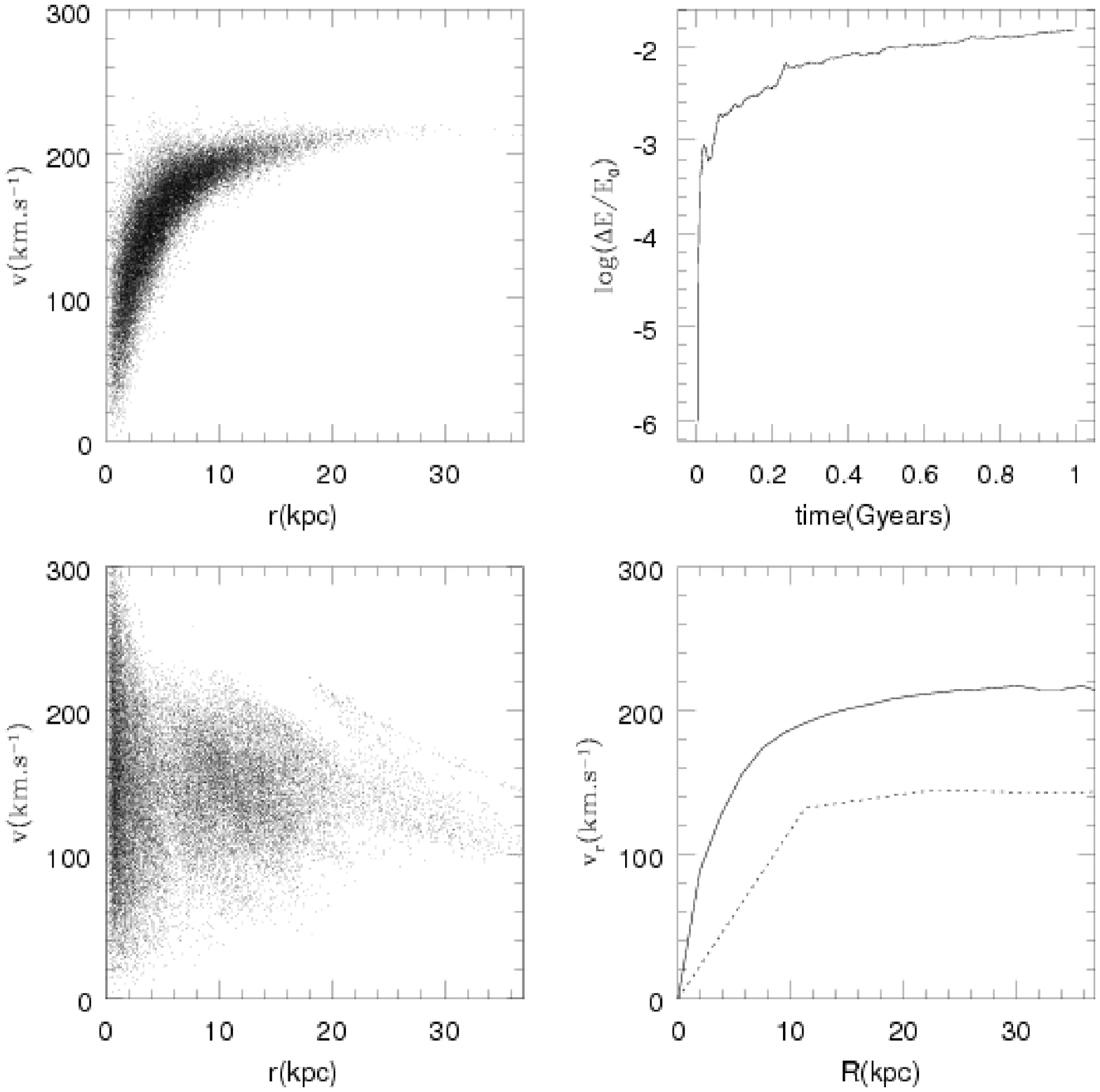}
\caption{Top left: phase space for initial snapshot data. Top right: energy conservation of the simulation. Bottom left: phase space for final snapshot data at 1 Gyr. Bottom right: rotation curves for initial (solid line) and final (dashed line) snapshots. \label{fig7}}
\end{figure}

\begin{figure}
\epsscale{1.0}
\plotone{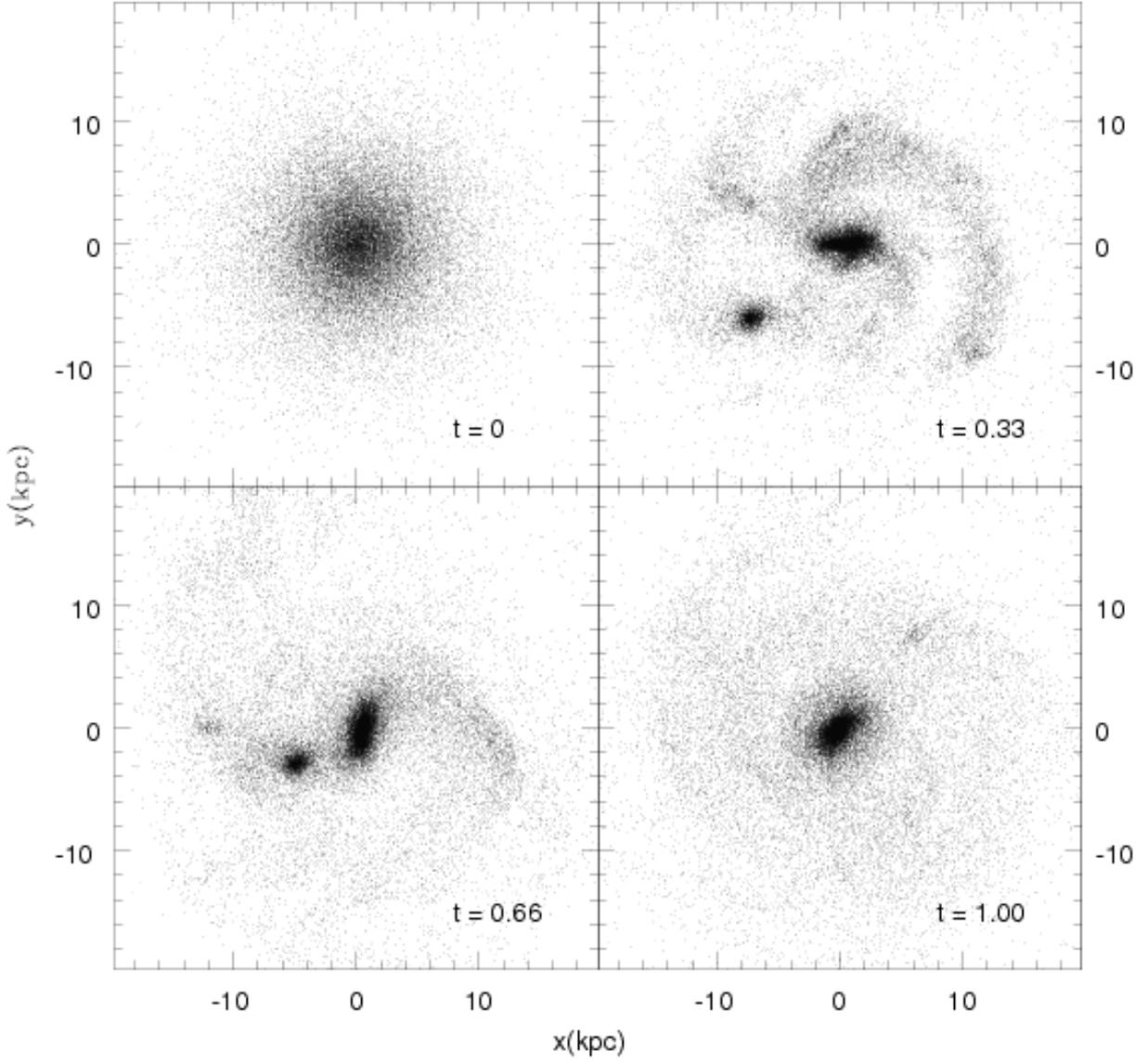}
\caption{BM disk at $z$=projection at 0, 0.33, 0.66, and 1 Gyr of simulated time (indicated in the respective boxes). \label{fig8}}
\end{figure}

\begin{figure}
\epsscale{1.0}
\plotone{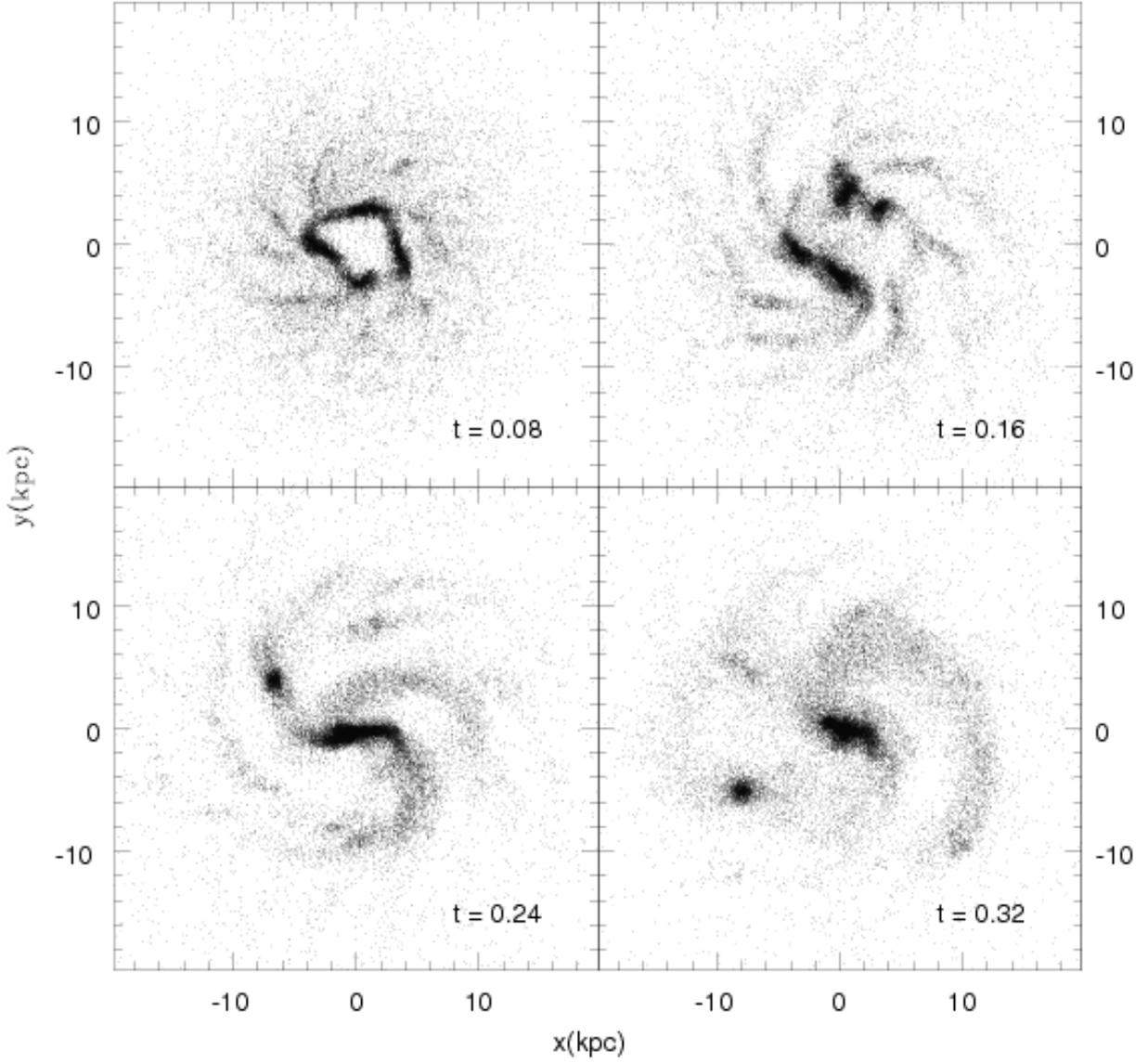}
\caption{First 300 Myr of simulated time to the MB disk at $z$-projection. Time is indicated in the respective boxes. \label{fig9}}
\end{figure}

It was just the results of model II  that lead us to make a pure Moffatian disk; our model III whose simulations were made with the same parameters used in the models I and II.  The results of model III can be seen in Figures \ref{fig7}--\ref{fig9}.

In the top right panel of Figure \ref{fig7}, one finds the phase-space points. These points are distributed somewhat differently as compared to previous models, because the rotation curve has now a different shape. Also, after 1 Gyr of simulated time, we note that the scattering in the phase space plane is more prominent than in the model II and the initial configuration is completely lost.

This effect is neither due to simulation inaccuracies nor to the resolution of the simulation, as can be seen that the energy violation (see the top right panel of Figure \ref{fig7}) is small, namely, $\log{\Delta E/ E_0} \lesssim -2.0$.

Our results show in the end that the exponential profile is not consistent with the Moffatian potential.  As occurred with model II, the rotation curve of the model III is below the initial one. For $\rm{R < 10}$ kpc, the rotation curve is linear and associated with the central core movement, similar to a rigid body rotation. For $\rm{R > 10}$ kpc, the rotation curve is similar to a typical rotation curve of late-type systems, but, in analogy with model II, this similarity is only apparent, as we will see below.

Figures \ref{fig8} and \ref{fig9} display the particle's positions at $(x,y)$ plane, where the simulated time is indicated in the respective boxes.  Note that, at initial times, the same atypical central void appears with its surrounding ring, that collapses and fragments in two small clumps, that evolves separately eventually merging at $t \sim 0.9$ Gyr.

This simulation shows that for a typical initial equilibrium configuration, BM's disk systems evolve in such way that we cannot produce models in agreement with the observed values, since atypical configurations appear.

The mixing in the phase space, for example, shows us that our BM's models do not recover the observational characteristics and, therefore, this model cannot be considered as a good representation of disk galaxies.

But, one could argue that the snapshot at time $t=1$ Gyr resembles a typical late-type system, and this model could be considered reliable, since after $t \gtrsim 0.9$ Gyr this configuration seems to be stable. Later on we consider this issue again by studying the density profile of the disk.

We now compare the rotation curve that comes out from our simulations with the one that
comes from the centrifugal equilibrium, i.e. $v^2(r)=r\, a(r)$, where $a(r)$ is given by Equation
(\ref{moffat_aceleration2}). This rotation curve reads

\begin{equation}
v(r) = \sqrt{ \frac{G_0 M(r)}{r}} \left\{ 1 + \sqrt{\frac{M_0}{M}} \left[1-\rm{exp}(-r/r_0)(1+r/r_0)\right] \right\}^{ \frac{1}{2}},
\label{rotacaolinear}
\end{equation}

\par\noindent where $M(r)$ is obtained from our simulation. Note that the above equation is just the one used by BM.

\begin{figure}
\epsscale{1.0}
\plotone{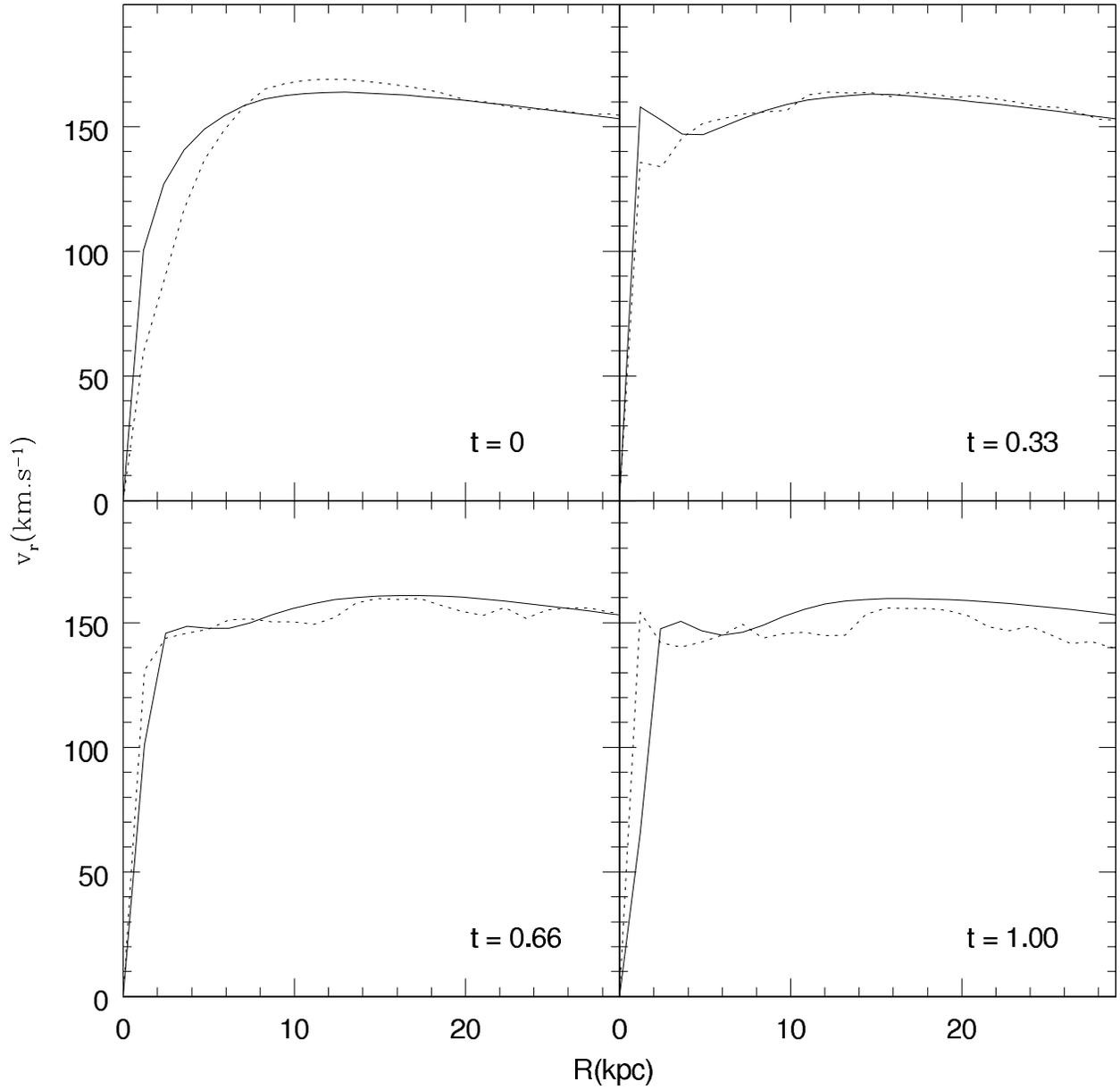}
\caption{Moffatian rotation curves (model III). Solid lines stand for the rotation curve given by Equation (\ref{rotacaolinear}) and dashed lines for the rotation curves obtained from the snapshots. Time is indicated
in the respective boxes.\label{fig10}}
\end{figure}

In Figure \ref{fig10}, we show the result of this comparison. Note the similarity between the rotation curves, even when considered different simulated times. The differences come from particle noise, very common in $N$-body samplings after evolving the system.  This result shows the consistency between our simulations and the calculation done by BM.

An important issue is to examine the radial density profile of the simulated disks. A reliable model of a spiral galaxy must have at the end of the simulation the same initial profile.

\begin{figure}
\epsscale{1.0}
\plotone{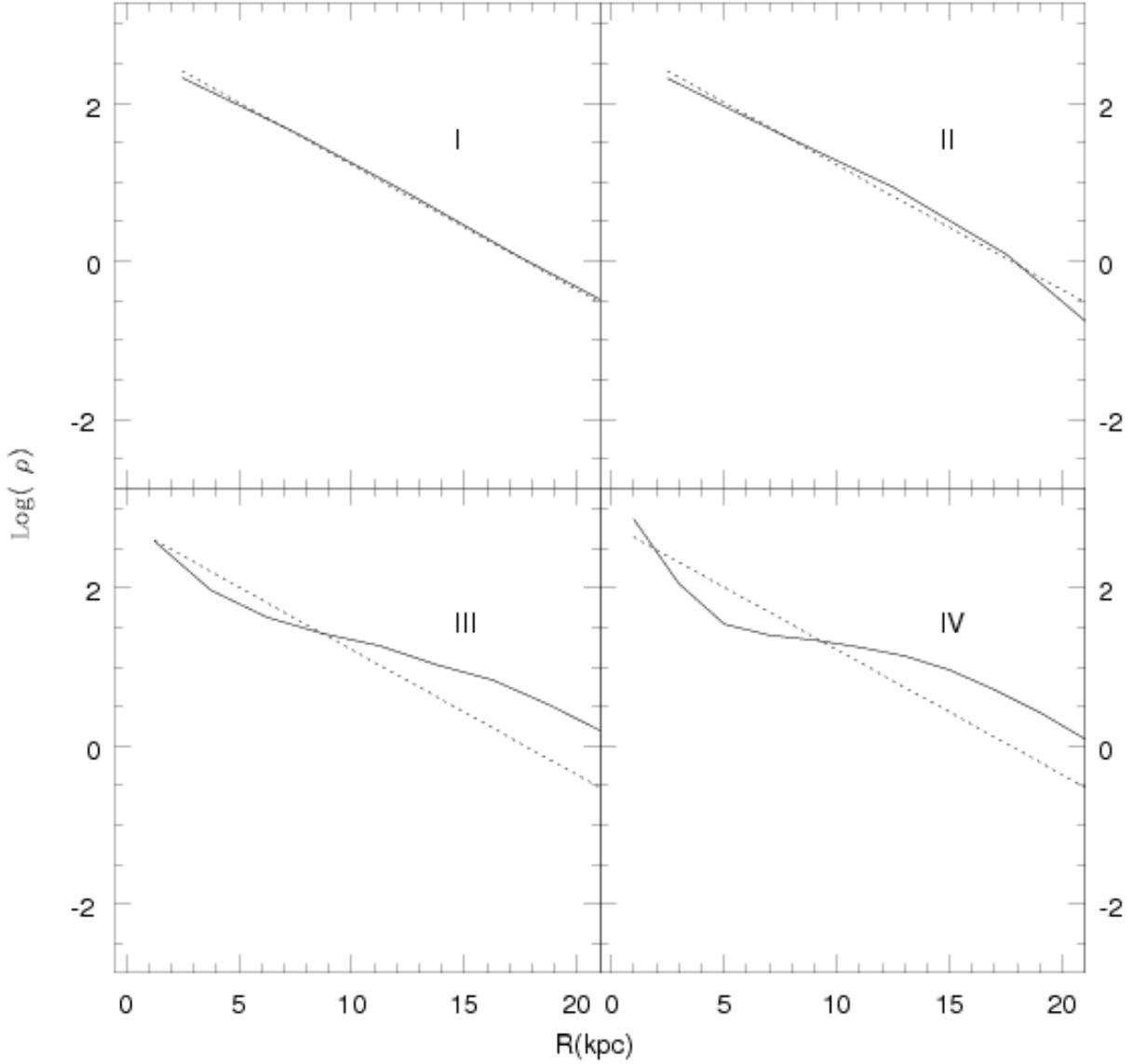}
\caption{Radial density profiles of the simulated models. Horizontal axis: radial distance $R$ from the center of gravity. Vertical axis shows the $\log$ of the particle counts, $\log(\rho)$, where $\rho$ is the number of particles per area unit.  Frame I shows the initial values of the density profile, represented by solid lines. Frame II: the final profile of the Newtonian model. Frame III: the profile of the pseudo-Moffatian disk, and Frame IV: the profile of the Moffatian disk. All frames display the exponential profile in dotted lines that match with the initial counts.  \label{fig11}}
\end{figure}

In Figure \ref{fig11}, the radial density profile of the simulated disks is calculated at $t = 0$ and $t=1$ Gyr. This figure presents the logarithmic of particle counts per unit of area, i.e., $\log(\rho)$,  in the $(x,y)$ plane, that was divided in many concentric rings, in cylindrical coordinate system. We have also estimated the initial exponential profile analytically, which is indicated by dotted lines.

Frame I indicates the Newtonian initial profile, estimated for all the particles and analytically both profiles coincide exactly. Frame II shows the final Newtonian profile; Frame III shows the final profile of the model II; and Frame IV shows the final BM profile (model III). Each frame has its initial exponential profile for comparison.  Note that the Newtonian disk maintains its initial profile, in spite of particle noise, very common in $N$-body samplings after evolving the system. This figure shows that bars and spiral arms left the exponential profile almost unchanged.

For the Frame III, however, we conclude that the simulations of the model II have minimized the central density and maximized the density at $R > 10$ kpc. In the phase space, this can be noted by the migration of particles to more distant regions, where their orbits are more stable, increasing the density there. The same happens with BM disks, as we can observe in Frame IV.

It is worth mentioning again that our BM disk has the same initial profile of the Newtonian model, and Frame I represents all initial profiles, due to the fact that all initial disks are modeled by the same exponential-Spitzer law. We note that, in spite of the final snapshot configurations of the models II and III mimic disks, their density profiles are very different from a canonical exponential-Spitzer law, from which they were built.  As a conclusion, the Moffatian potential cannot maintain exponential-Spitzer disks in dynamical equilibrium and therefore cannot be consistent with observations.

\begin{figure}
\epsscale{1.0}
\plotone{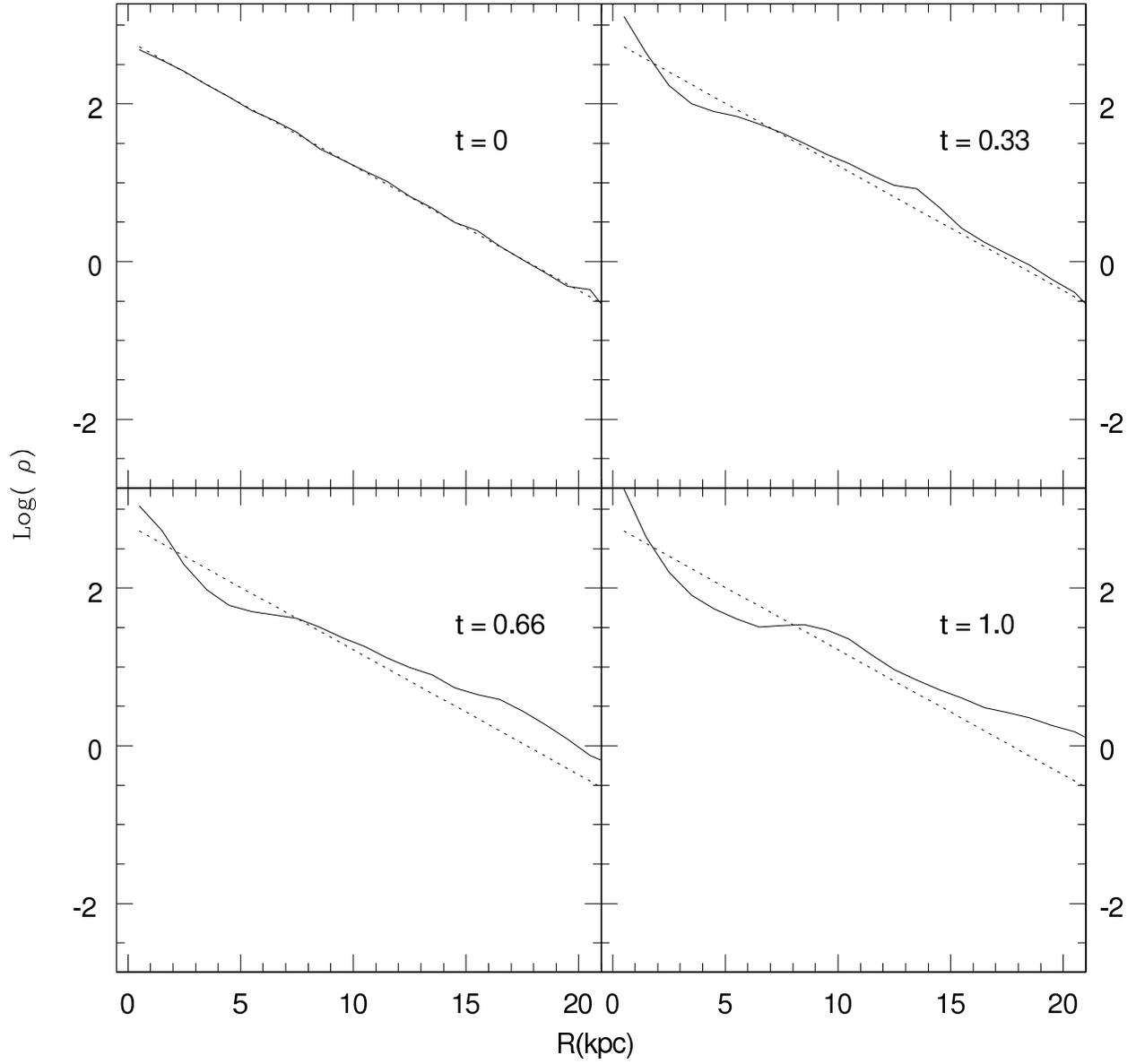}
\caption{Evolution of the radial density profiles of the model III. Time is indicated in the respective boxes. \label{fig12}}
\end{figure}

In Figure \ref{fig12}, we present, for model III, the radial density profile for different simulated times.
This figure clearly shows that, even starting from a disk in equilibrium, the Moffatian gravity does not maintain the exponential-Spitzer disk with the same initial density profile. As time goes on, the profile turns gradually and is definitely different from the exponential profile. This shows that simulating models are a more robust tool than simply adjust profiles in static models, because the evolution of the system cannot be followed in this last approach.

In conclusion, an alternative gravitational law must explain not only the rotation curves of spiral galaxies,
but also their density profiles. In this way, simulations can be a powerful tool to deal with
such an issue.

\subsubsection{An additional test}

Note that depending on the values of $r_0$ and  $M_0$, the Newtonian gravitation would be naturally recovered.
Therefore, as an additional test we examine such an issue, which could help one to see how the structure of
a disk galaxy would be modified by an alternative gravity law.

We start from a Moffatian disk, just the one present, for example, in the first snapshot of Figure \ref{fig8}, whose initial rotation curve is shown in Figure \ref{fig7}. Then, we set $r_0 = 1000 kpc$ and $M_0=1 \times 10^{10}$ $M_{\odot}$ and follow how the disk evolves.

\begin{figure}
\epsscale{1.0}
\plotone{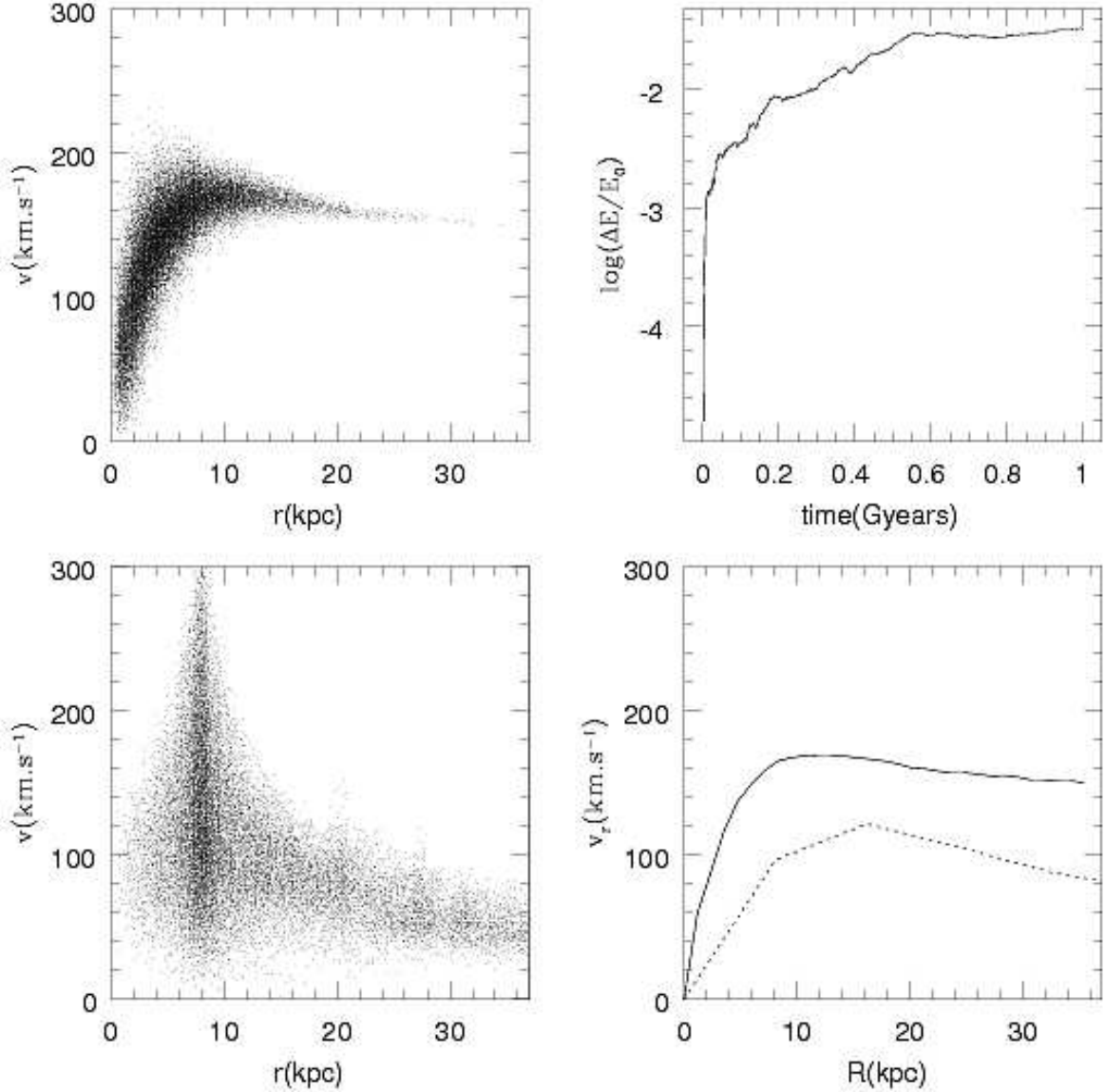}
\caption{Model III with the same initial snapshot used in Figure \ref{fig8}, obtained with the Gadget-2 modified code in the Newtonian limit, i.e., with $r_0 = 1000$ kpc and $M_0=1 \times 10^{10}M_{\odot}$. Top left: phase space for initial snapshot data. Top right: energy conservation of the simulation. Bottom left: phase space for final snapshot data at 1 Gyr. Bottom right: rotation curves for initial (solid line) and final (dashed line) snapshots. \label{fig13}}
\end{figure}

\begin{figure}
\epsscale{1.0}
\plotone{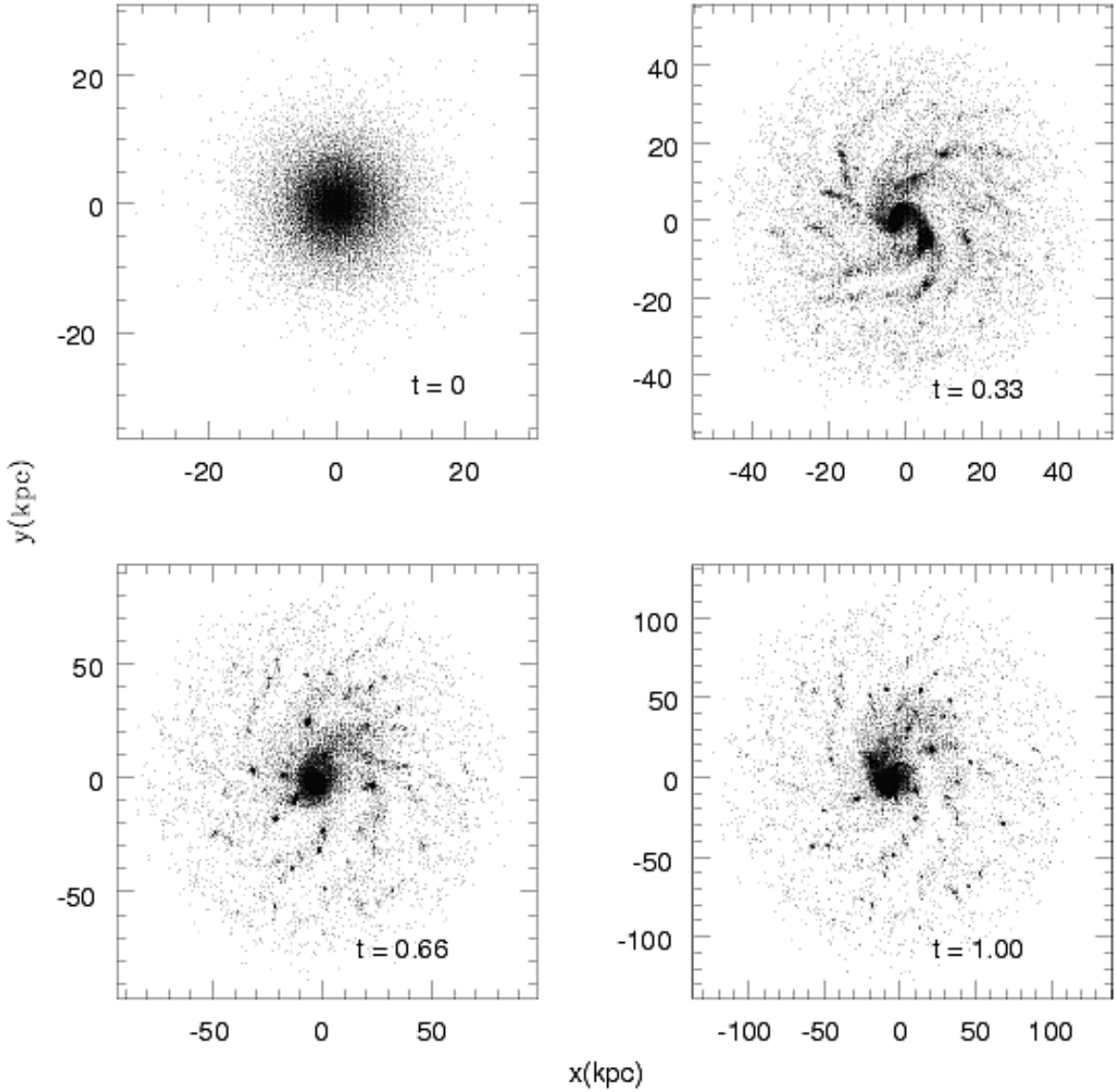}
\caption{BM disk at $z$-projection at 0, 0.33, 0.66, and 1 Gyr of simulated time (indicated in the respective boxes) obtained with the Gadget-2 modified code in the Newtonian limit, i.e., with $r_0 = 1000$ kpc and $M_0=1 \times 10^{10}M_{\odot}$. \label{fig14}}
\end{figure}

The results of this simulation are shown in Figure \ref{fig13} and \ref{fig14}. For comparison,
we also performed simulations using the Newtonian Gadget-2. Both the calculations produce identical
results. Therefore, the Newtonian calculation is recovered.

In particular, note in Figure \ref{fig13} the rotation curve after 1 Gyr of simulated time; for large values of $R$, the velocity is a decreasing function of $R$, as expected for a Newtonian disk.

It is also interesting to compare the Moffatian disk snapshots and Newtonian ones. Comparing Figures \ref{fig8} and \ref{fig14}, one clearly sees how different these configurations are.

\section{Conclusions}
\label{conclusao}

Studies of alternative theories of gravity are usually made statically, instead of modeling ``live'' systems. For example, in the study of spiral galaxies centrifugal equilibrium is usually considered. The problem with this approach is that the secular evolution of these systems cannot be followed.

In the present paper, in particular, we follow the techniques used by \citet{ca2009a,ca2009b,ca2009c} and probe the BM's model, verifying if the Moffatian gravity is dynamically consistent to account for spiral galaxies. A dynamical study is a robust approach, since it considers the distribution functions and Jean's equations that can maintain $N$-body systems in equilibrium.

We performed the simulations with a modified and tested version of the Gadget-2 code. We replaced the Newtonian potential and acceleration by the expressions given by Equations (\ref{moffat_equation2}) and (\ref{moffat_aceleration2}) just in the respective code's instructions. We have extensively  tested the code efficiency to calculate potentials via the tree method, using a typical initial snapshot composed by a disk embedded in a dark matter halo.

From our simulations, it follows that the Moffatian potential cannot generate exponential disks in equilibrium, although the rotation curves are nearly flat throughout the whole simulated time.  The final (stable) equilibrium configuration is very different from the exponential profile, as we have explained above.  So, if we consider that exponential profiles are reliable, as it seems to be the case, since they are consistent with observations, it is hard to believe that BM model is better than the Newtonian model (baryonic disk embedded in a dark matter halo) to account for the spiral galaxy structure. Recall that \citet{bm2006a} claim  that $\rm{M_0}$ and $\rm{r_0}$ fit very well many (observed) rotation curves, but they have only analyzed static models (centrifugal equilibrium configurations).

Finally, it is worth mentioning that it would be very interesting to simulate the BM potential for standard elliptical systems and to probe its influence over their global properties, like we have done in this work. A forthcoming paper will consider such an issue.


\acknowledgments
We are grateful to the Brazilian agencies CAPES, CNPq and Fapesp for support. We also thank Nick Gnedin for providing us a visualization software. Last but not least, we thank the referee for useful suggestions and criticisms, which greatly improved the final version of our paper.

\clearpage

\clearpage

\clearpage

\clearpage

\clearpage

\clearpage

\clearpage

\clearpage

\clearpage

\clearpage

\clearpage

\clearpage

\clearpage

\clearpage

\end{document}